\documentclass[12pt]{article}
\usepackage{amsmath}
\usepackage{graphicx}
\usepackage{enumerate}
\usepackage{natbib}
\usepackage{url} % not crucial - just used below for the URL 

\usepackage{amssymb}
\usepackage{bm}

\usepackage{subcaption}
\newcommand{\blind}{1}

\newcommand{\Rset}{\mathbb{R}}

\newcommand{\ydot}{{\dot{y}}_{dt}}

\newcommand{\dif}{\mathop{}\!\mathrm{d}}
 
\begin{document}

\def\spacingset#1{\renewcommand{\baselinestretch}%
{#1}\small\normalsize} \spacingset{1}

%%%%%%%%%%%%%%%%%%%%%%%%%%%%%%%%%%%%%%%%%%%%%%%%%%%%%%%%%%%%%%%%%%%%%

\if1\blind
{
  \title{\bf Conex-Connect: Learning Patterns in Extremal Brain Connectivity From Multi-Channel EEG Data}
  \author{{\bf{Matheus B Guerrero, Rapha{\"e}l Huser, and Hernando Ombao}}\vspace{.1cm}\\
    \small{King Abdullah University of Science and Technology (KAUST)} \vspace*{-.1cm} \\
    \small{Statistics Program, CEMSE Divison}}
    
  \maketitle
} \fi

\if0\blind
{
  \bigskip
  \begin{center}
    {\LARGE\bf Conex-Connect: Learning Patterns in Extremal Brain Connectivity From Multi-Channel EEG Data}
\end{center}
  \medskip
} \fi

\bigskip\vspace*{-.5cm}
\begin{abstract}
\noindent Epilepsy is a chronic neurological disorder affecting more than 50 million people globally. An epileptic seizure acts like a temporary shock to the neuronal system, disrupting normal electrical activity in the brain. Epilepsy is frequently diagnosed with electroencephalograms (EEGs). Current methods study the time-varying spectra and coherence but do not directly model changes in extreme behavior. Thus, we propose a new approach to characterize brain connectivity based on the joint tail behavior of the EEGs. Our proposed method, the conditional extremal dependence for brain connectivity (Conex-Connect), is a pioneering approach that links the association between extreme values of higher oscillations at a reference channel with the other brain network channels. Using the Conex-Connect method, we discover changes in the extremal dependence driven by the activity at the foci of the epileptic seizure. Our model-based approach reveals that, pre-seizure, the dependence is notably stable for all channels when conditioning on extreme values of the focal seizure area. Post-seizure, by contrast, the dependence between channels is weaker, and dependence patterns are more ``chaotic''. Moreover, in terms of spectral decomposition, we find that high values of the high-frequency Gamma-band are the most relevant features to explain the conditional extremal dependence of brain connectivity.
\end{abstract}

\noindent%
\vfill

\newpage
\spacingset{1.5} % DON'T change the spacing!

% ---- ------- ---- % ------- % ---- ------- ---- %
% ---- SECTION ---- % SECTION % ---- SECTION ---- %
% ---- ------- ---- % ------- % ---- ------- ---- %
\section{Introduction}\label{chap:intro}

\begin{figure}[!htp]
\centerline{
\includegraphics[width=\textwidth]{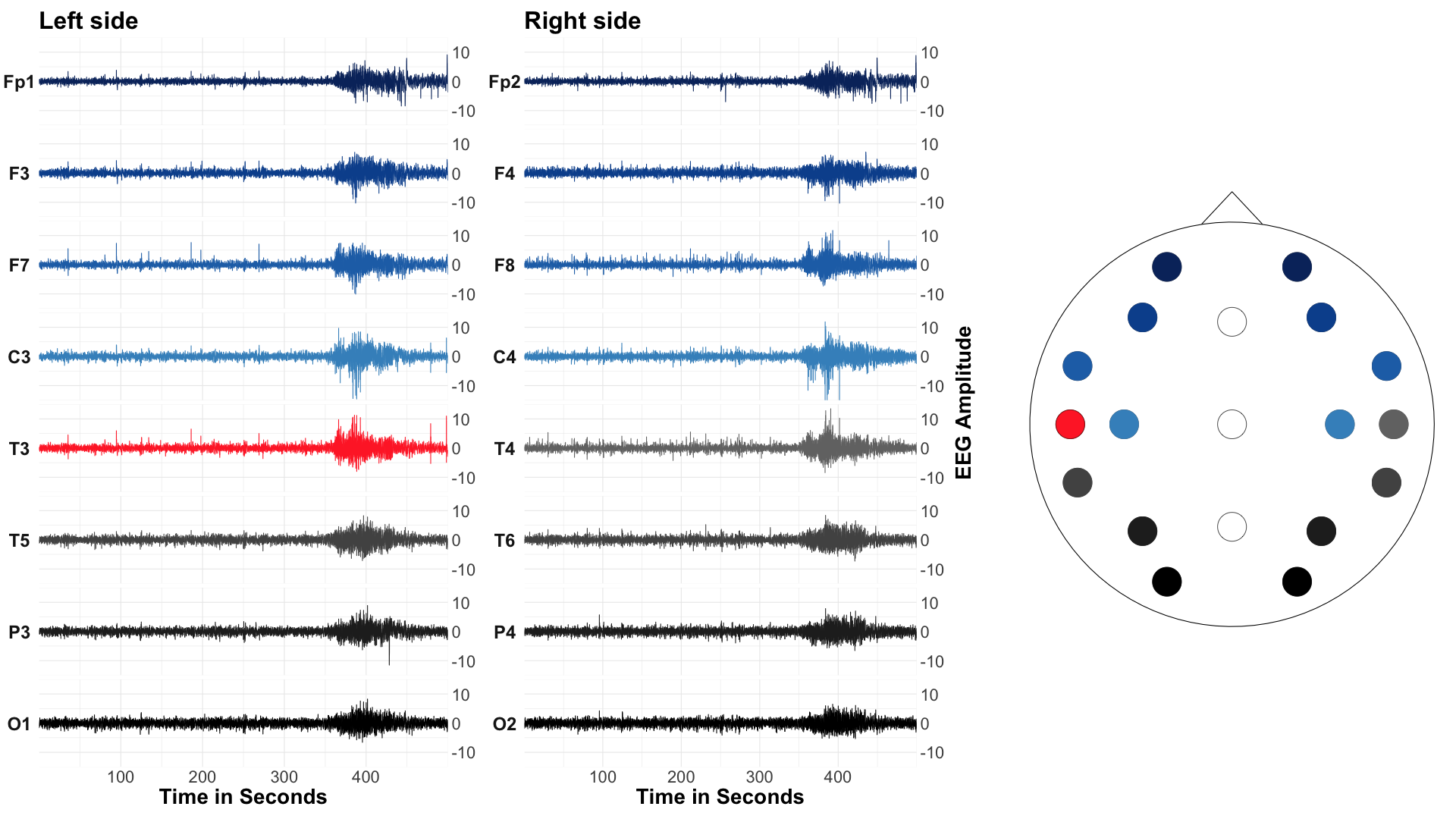}
}
\caption{Recorded EEG signals ($10$--$20$ electrode system) from a female patient diagnosed with left temporal lobe epilepsy during an epileptic seizure. The seizure onset happens roughly at $t = 350$ seconds on the reference channel T3 (red). The dataset consists of $50,000$ time points at a sampling rate of $100$ Hz. The central temporal (Fz), central (Cz) and 
central parietal (Pz) channels are not displayed in the figure, albeit they are included in the analysis.}
\label{fig:eeg_scalp}
\end{figure}

Electroencephalograms (EEGs) are multidimensional spatio-temporal signals that measure brain electrical activity from electrodes placed on the scalp. EEGs capture changes in brain signals following a shock to the neuronal system, such as an external stimulus, stroke, or epilepsy. These shocks show a profound impact on the neuronal system, including changes in frequency content, wave amplitudes and connectivity structure of the network of signals. Often these shocks have an impact on the distributions (in particular, at the tails). Here, our goal is to develop a new statistical approach for investigating changes in the extremal dependence structure, i.e., the dependence structure prevailing in the tail of the distribution, between EEG channels during an epileptic seizure. While classical methods mostly rely on the behavior in the bulk (or around the center) of the distribution, our extreme-value method is natural and theoretically justified for modeling extremely large signal amplitudes that are observed during the onset of an epileptic seizure. Therefore, the proposed method, which provides new insights into ``extremal'' brain connectivity, may have a potential impact on public health, given that epilepsy affects nearly 50 million people worldwide \citep{who19}. The anticipated impact will be deeper understanding on the etiology of seizure and refined diagnosis brought about by the ability to differentiate between seizure subtypes and hence develop more targeted treatments. Figure \ref{fig:eeg_scalp} displays EEG traces of the left and right sides of the brain during an epileptic seizure from a patient previously diagnosed with left temporal lobe epilepsy. The seizure onset is believed to be at channel T3 (in red).

The most frequently used measure of dependence is cross-correlation. Consider a bivariate random vector $(X,Y)^{T}$ with joint density $f(x,y)$ with support on the domain $\mathcal{D}$ with marginal means $\text{E}(X) = \text{E}(Y) = 0$ and variances $\text{var}(X) = \text{var}(Y) = 1$. Then, the cross-correlation between $X$ and $Y$ is $\rho_{XY} = \text{E}(XY) = \int_{\mathcal{D}} xy f(x, y) \dif x \dif y$, where the integral is calculated across the entire domain $\mathcal{D}$. Indeed, classical time-domain and frequency-domain measures of dependence, in multivariate time series, are all derived by some averaging across the entire domain $\mathcal{D}$. The main limitation of these measures is that they cannot capture local, and sometimes conflicting, variations in the dependence structure that may happen across the domain $\mathcal{D}$. That is, the dependence structure at the ``center'' of the distribution might be different from that at the ``tails.'' This is especially important when analyzing EEGs that are recorded during an epileptic seizure when there is an abnormal disruption in electrical activity in the brain that results in increased amplitude and changes in spectral decomposition at the localized focal region. In this paper, we develop a new statistical procedure for studying how an extreme signal amplitude caused by an epileptic seizure at the focal region (left temporal lobe T3) may trigger a change in the signal amplitude of another channel of the brain network. In other words, the proposed method will be utilized to study the conditional extremal brain connectivity of an epileptic person. 

Our proposed method, conditional extremal dependence for brain connectivity (\textbf{Conex-Connect}, in short), jointly utilizes extreme-value theory and spectral analysis to model the conditional extremal dependence of brain connectivity. Conex-Connect adapts the conditional extremes model \citep{HT04} for multivariate extremes to time-varying brain signals and further extends it to the spectral setting where we study the impact of extreme oscillations in one channel on the behavior of other channels; see \cite{davison15} for a general overview of the extreme-value theory.

Consider a random vector $(X,Y)^{T}$ with marginal cumulative distribution functions (CDFs) $F_X$ and $F_Y$, respectively. The Heffernan and Tawn (H\&T) model is an asymptotically motivated model for the conditional distribution of $X$ given that $Y$ is large, i.e., $X \mid Y > F_{Y}^{-1}(p)$, as $p \uparrow 1$. In finite samples, we condition on $Y$ being larger than its $p$-quantile, for some (fixed) high non-exceedance probability $p \in (0, 1)$. The H\&T model can capture the two possible regimes of extremal dependence between $X$ and $Y$: asymptotic independence, which arises when $\chi = \lim_{p \to 1} \Pr \left(X > F_{X}^{-1}(p) \mid Y > F_{Y}^{-1}(p)\right) = 0$, and asymptotic dependence, which arises when $\chi > 0$. The distinction between these two regimes is key for extrapolating to higher levels of the conditioning variable $Y$. However, at the onset of a seizure, EEGs display sudden bursts or increases in amplitude, which may be attributed to some (but not all) frequency bands. One limitation of the H\&T model is that it is unable to distinguish such features and hence needs to be adapted accordingly. Our proposed Conex-Connect method overcomes these limitations: it examines extremal dependence behavior in oscillatory components of brain signals for specific frequency bands.

Moreover, the H\&T model is designed for stationary data with respect to covariates. Thus, it needs to be adapted to seizure EEG signals which are highly non-stationary -- both in their marginal and dependence behavior. Here, we further modify the H\&T model to flexibly capture the time dynamics of extremal brain connectivity based on an improved version of the penalized piecewise constant (PPC) approach of \cite{Ross18}.

The remainder of the paper is organized as follows. Section \ref{chap:method} presents our proposed extreme-value approach to analyze EEG data. Section \ref{chap:application} illustrates our novel methodology by an application to a real dataset from a patient with left temporal lobe epilepsy. Section \ref{chap:conclusion} concludes and discusses perspectives on future research.

% ---- ------- ---- % ------- % ---- ------- ---- %
% ---- SECTION ---- % SECTION % ---- SECTION ---- %
% ---- ------- ---- % ------- % ---- ------- ---- %
\section{Conditional extremal dependence for connectivity (Conex-Connect)}\label{chap:method}

% ---- SUBSECTION ---- %
\subsection{General setting and method overview}

Let $\dot{Y}_{dt}$ be the EEG amplitude, in absolute value, of channel $d \in \{1, \ldots, D\}$ at time $t \in \{1, \ldots, T\}$. Alternatively, it may also represent the amplitude of filtered signal component at a given frequency band; see Section \ref{chap:spectral} for more details on spectral decomposition. Figure \ref{fig:eeg_scalp} displays the behavior of these multi-channel time series. In a nutshell, our proposed method proceeds as follows. The first step is to select a reference channel (i.e., the \textit{conditioning variable}). In our case, it is natural to select the left temporal channel T3 (in red; Figure \ref{fig:eeg_scalp}) as the reference channel because it is believed to be the focal point of seizure discharges for this particular patient. We label it as  $\dot{Y}_1=\{\dot{Y}_{1t}\}_{t=1}^T$. The remaining channels are called the associated channels (\textit{associated variables}) and are labeled as $\{\dot{Y}_{d}\}_{d=2}^{D}$, where $\dot{Y}_d=\{\dot{Y}_{dt}\}_{t=1}^T$. Next, since the dependence structure in the brain network is thought to evolve across time, we segment the multi-channel EEG dataset into $B$ (approximately) time-homogeneous blocks. Observations within the same time block are assumed to have common extremal characteristics. Finally, we model the changes in both the marginal and dependence properties.

More precisely, for each channel $d \in \{1, \ldots, D\}$, we fit a PPC marginal model by assuming that high threshold exceedances follow a generalized Pareto (GP) distribution with a time-varying scale parameter $\nu_{db} > 0$, $b \in \{1, \ldots, B\}$. Using the empirical CDF below the threshold, we then transform the data from their original scale to a common marginal scale chosen to be the Laplace distribution, based on the probability integral transform. The original data are denoted by $\dot{Y}_d$, while the transformed Laplace-scale data are denoted by $Y_d$, $d\in\{1,\ldots,D\}$. The GP marginals are transformed to the same standard scale so that they become comparable. Moreover, the Laplace distribution is chosen because the H\&T model expressed on this scale yields a wide range of extremal dependence structures, from asymptotic independence with negative association to asymptotic dependence with positive association \citep{Keef13}.

To estimate the conditional extremal dependence of $Y_{d}$ given that $Y_{1}$ exceeds a high threshold, we finally fit the H\&T model with a time-varying dependence parameter $\alpha_{db} \in [-1, 1]$, which captures the evolution (over time) of the strength of the linear dependence in the joint tails. The specific elements of the H\&T model, including the parameter $\alpha_{db}$, are discussed in Section \ref{chap:dependence}. The penalized term in the PPC approach stems from roughness-penalization parameters included in the likelihood functions to control the extent of variation of the time-varying parameter estimates for the marginal and dependence models. We assess the uncertainty of the estimates through a bootstrap technique that accounts for the temporal dependence present in the data. The optimal penalty parameter is obtained using a walk-forward cross-validation procedure designed to keep the temporal dependence in the data. The next subsections provide details on the two stages of the Conex-Connect method: the marginal modeling of EEG channels and their connectivity characterization.

% ---- SUBSECTION ---- %
\subsection{Marginal model}\label{chap:marginals}

For each channel $\dot{Y}_{d} = \{\dot{Y}_{dt}\}_{t=1}^{T}$, $d \in \{1,\ldots,D\}$, and for some high non-exceedance probability $\tau_{d} \in \mathcal{T}_d \subset (0, 1)$, we define $\psi_{db}(\tau_{d})$ as the empirical $\tau_d$-quantile for the $b$-th time block ($b=1,\ldots, B$). Then, motivated by extreme-value theory, we model the upper tail of $\dot{Y}_{dt}$ using the GP distribution, i.e.,
$$
F_{GP}\left(\dot{y}_{dt}; \xi_{d}, \nu_{db}, \psi_{db}(\tau_{d})\right) = 
\Pr\left(\dot{Y}_{dt}\leq\dot{y}_{dt} \mid \dot{Y}_{dt}>\psi_{db}(\tau_{d})\right) =
1 - \left[1 + \frac{\xi_{d}}{\nu_{db}}\left\{\dot{y}_{dt} - \psi_{db}(\tau_{d})\right\}\right]^{-\frac{1}{\xi_{d}}},
$$
for $\dot{y}_{dt} \in \left(\psi_{db}(\tau_{d}), \dot{y}_{db}^{+}\right]$, where $\dot{y}_{db}^{+} = \psi_{db}(\tau_{d}) - \nu_{db}/\xi_{d}$ if $\xi_{d}<0$, and  $\dot{y}_{db}^{+}=\infty$, otherwise. Here, $\xi_{d} \in \Rset$ is the shape parameter, assumed constant across time blocks, while $\{\nu_{db}\}_{b=1}^{B} \in (0,\infty)^{B}$ are the block-wise scale parameters. In the above expression, it is implicitly understood that time $t$ is contained within the $b$-th time block, but in practice we need to identify the correct time block for each time point, and assign parameter values accordingly. For each channel $d$, the shape parameter $\xi_d$ controls the heaviness of the tail of the GP distribution (when compared to the tail of an exponential distribution), and we here keep it time-constant for reasons of parsimony and because this parameter is usually difficult to estimate. Depending on the value of $\xi_d$, there are three types of upper tail: bounded ($\xi_d < 0$), light ($\xi_d = 0$), and heavy ($\xi_d > 0$). As special cases, the GP distribution contains the uniform distribution ($\xi_d = -1$), the exponential ($\xi_d = 0$), and the (shifted) Pareto ($\xi_d >0$).

Note that each channel has its own non-exceedance probability $\tau_{d}$, indicating the flexibility of the model in capturing the channel-specific extremal characteristics. In addition, despite $\tau_{d}$ being invariant over time blocks, it does not imply threshold invariance (over time) since the $\tau_{d}$-quantile is specific to each time block.

Let $\bm{\theta}_{d} := \left(\{\nu_{db}\}_{b=1}^{B},\xi_{d}\right)^{T} \in \bm{\Theta} := (0,\infty)^B \times \Rset$ be the parameter vector of the marginal GP model for channel $d$. Under the working assumption of independence, the likelihood function is
$$\mathcal{L}_{\tau_{d}}(\bm{\theta}_{d}) = \prod\limits_{b=1}^{B}\prod\limits_{\substack{t \in T_b \\ \ydot > \psi_{db}(\tau_{d})}}\dfrac{1}{\nu_{db}}\left[1 + \dfrac{\xi_{d}}{\nu_{db}}\left\{\ydot - \psi_{db}(\tau_{d})\right\}\right]^{-\frac{1}{\xi_{d}} - 1},
$$
where $T_{b}$ is the set of time points within block $b$, and $\{\dot{y}_{dt}\}_{t=1}^T$ are the observed data. To control the variance of temporal fluctuations in the estimated GP scale parameters, $\nu_{db}$, we add a roughness-penalization parameter $\lambda_{d} > 0$. The negative penalized log-likelihood is
\begin{equation}
\ell_{\tau_{d}, \lambda_{d}}(\bm{\theta}_{d}) = -\log{\mathcal{L}_{\tau_{d}}(\bm{\theta}_{d})} + \lambda_{d}\left\{\dfrac{1}{B}\sum\limits_{b=1}^{B}\nu_{db}^{2}-\left(\dfrac{1}{B}\sum\limits_{b=1}^{B}\nu_{db}\right)^{2}\right\}.
\label{eq:gploglik}
\end{equation}
Marginal parameter estimates are obtained by minimizing \eqref{eq:gploglik}, while $\lambda_d$ is selected by cross-validation; see Section \ref{chap:diag}. Note that larger $\lambda_d$ values imply an increased smoothness of estimates of the GP scale parameter $\nu_{db}$ across time blocks. Also, given $\lambda_d$, each channel has a different penalized log-likelihood, i.e., each marginal fit has a different set of parameters, consisting of $(B+1)$ parameters from the GP fit.

Now, using the probability integral transform, we transform data to the standard Laplace scale. For observations below the threshold $\psi_{db}(\tau_{d})$, we use the empirical CDF, denoted here by $F_{E}(\cdot)$. First, the raw data, $\{\dot{y}_{dt}\}_{t=1}^{T}$, is transformed to the uniform scale, $\{u_{dt}\}_{t=1}^{T}$, as follows: $u_{dt} = \tau_d F_{E}(\dot{y}_{dt})$, if $\dot{y}_{dt} \leq \psi_{db}(\tau_{d})$; and $u_{dt} =\tau_d + (1-\tau_d) F_{GP}(\dot{y}_{dt})$, if $\dot{y}_{dt} > \psi_{db}(\tau_{d})$. Then, we use the inverse of the standard Laplace CDF, $F_{L}(\cdot)$, i.e., $y_{dt} = F_{L}^{-1}(u_{dt})=\text{sign}(0.5 - u_{dt})\log{\left(2\min{\{1-u_{dt}, u_{dt}\}}\right)}$, to obtain common standard Laplace margins, $\{y_{dt}\}_{t=1}^{T}$.

% ---- SUBSECTION ---- %
\subsection{Conditional extremal dependence model}\label{chap:dependence}

After fitting the marginal models for all $D$ channels, we obtain a standard Laplace-scale sample $\{y_{1t}, y_{2t},\ldots, y_{Dt} \}_{t=1}^{T}$, with $y_{1t}$ representing the transformed time series of the reference channel; here, T3. Here, the lowercase $\{y_{1t},\ldots,y_{Dt}\}_{t=1}^T$ denote (transformed) realized values, while the uppercase $\{Y_{1t},\ldots,Y_{Dt}\}_{t=1}^T$ denote the corresponding random variables. The next goal is to study the conditional dependence of the associated variables, $\{Y_{2t},\ldots, Y_{Dt} \}$, given that the conditioning variable, $\{Y_{1t}\}$, takes on a large value (in the upper tail of the distribution). Define $\tilde{\tau} \in \tilde{\mathcal{T}} \subset (0,1)$ to be a non-exceedance probability such that $\phi(\tilde{\tau})$ is the $\tilde{\tau}$-quantile of the standard Laplace distribution for the reference channel $Y_1$. Let $\tilde{\bm{\theta}}=\left(\{\alpha_{db}\}_{d=2,b=1}^{D, B}, \{\beta_{d}\}_{d=2}^{D}, \{\mu_{d}\}_{d=2}^{D}, \{\sigma_{d}\}_{d=2}^{D}\right)^{T} \in \tilde{\bm{\Theta}}:=[-1,1]^{(D-1)B}\times (-\infty,1]^{(D-1)}\times \Rset^{(D-1)}\times (0,\infty)^{(D-1)}$ be the parameter vector of the H\&T model for all channels. Thus, according to the H\&T model, for all $t\in T_b$ (within time block $b$) such that $y_{1t} > \phi(\tilde{\tau})$, conditional on $Y_{1t} = y_{1t}$, $Y_{dt}$ may be expressed as
\begin{equation}
Y_{dt} = \alpha_{db} y_{1t} + y_{1t}^{\beta_d} W_{dt},\quad d = 2, \ldots, D, \;\; b = 1, \ldots, B,
\label{eq:ht}
\end{equation}
where, for model estimation purposes, the components of the random variable $W_{dt}$ are assumed to be mutually independent and normally distributed with mean $\mu_{d}$ and standard deviation $\sigma_{d}$, both time-constant.

The parameters $\{\alpha_{db}\}_{d=2,b=1}^{D, B}$ are the ``first-order'' dependence parameters, and they capture the extent of block-wise linear extremal dependence between channels $Y_{d}$ and large $Y_{1}$, which is allowed to change over time (blocks). When $\alpha_{db} \in (0, 1]$, there is a positive linear association between $Y_d$ and large $Y_1$ within time block $b$. This association becomes stronger as $\alpha_{db}$ increases, with (positive) asymptotic dependence  corresponding to $\alpha_{db}=1$ and $\beta_d=0$. For $\alpha_{db} \in [-1, 0)$, the association is negative linear, and becomes stronger as $\alpha_{db} \to -1$. As $\alpha_{db} \to 0$, the linear dependence weakens. In particular, when $Y_d$ and large $Y_1$ are independent, then $\alpha_{db} =\beta_{d}=0$. On the other hand, the time-constant parameters $\{\beta_{d}\}_{d=2}^{D}$ may be considered as capturing the ``second-order'' dependence characteristics since they specify the spread of the data around the linear relationship given by $\{\alpha_{db}\}$ for increasing values of $Y_1$. As $\beta_d \to -\infty$, the distribution of the data (around the linear relationship dictated by the $\alpha$'s) becomes tighter for higher values of the conditioning variable. If $\beta_d > 0$, we have the opposite behavior; the distribution of the data has a wider spread around the linear relationship (given by the $\alpha$'s) for higher values of the conditioning variable. Notice that among all dependence parameters, $\alpha_{db}$ is the only one that is allowed to vary across time-blocks $b=1,\ldots,B$. The other parameters ($\beta_d$, $\mu_d$, $\sigma_d$) are intentionally kept constant over time in order to reduce the overall uncertainty, and avoid interferences with the estimation of $\alpha_{db}$, which drives the main dependence feature. 

From \eqref{eq:ht}, we may rewrite the model as
\begin{equation}
Y_{dt} \mid Y_{1t} = y_{1t} \sim \text{N}
\left(
\alpha_{db} y_{1t} + \mu_{d} y_{1t}^{\beta_d},\,\, 
(\sigma_{d} y_{1t}^{\beta_d})^2
\right),
\label{eq:w}
\end{equation}
for all $t\in T_b$ (within time block $b$) such that $y_{1t} > \phi(\tilde{\tau})$, $d = 2$, $\ldots$, $D$, and $b = 1$, $\ldots$, $B$. The parameters for each channel are estimated by minimizing the negative log-likelihood based on \eqref{eq:w}, i.e.,
\begin{equation}
\tilde{\ell}_{\tilde{\tau},d}(\tilde{\bm{\theta}}) = 
\sum\limits_{b=1}^{B}\sum\limits_{\substack{t \in T_b \\ y_{1t}>\phi(\tilde{\tau})}}
\left\{
\frac{1}{2}\log{(2\pi)} + \log{(\sigma_{d} y_{1t}^{\beta_d})}+ \dfrac{1}{2}\left( \frac{y_{dt} - (\alpha_{db} y_{1t} + \mu_{d} y_{1t}^{\beta_d})}{\sigma_{d} y_{1t}^{\beta_d}}\right)^{2}
\right\}.
\label{eq:loglik_channel}
\end{equation}
Here, there are ($B + 3$) parameters per channel $d$. \cite{Keef13} proposed additional constraints on $\tilde{\bm{\Theta}}$, leading to a smaller set of feasible parameters, which reduce the variance of the estimators and overcome complications on the modeling of negatively associated variables and parameter identifiability. Also, these additional constraints avoid drawing conditional inferences inconsistent with the marginal distributions. Here, we use a pragmatic approach and restrict $\{\beta_{d}\}_{d=2}^{D}$ to the interval $[0,1]$, to stabilize estimation and avoid unrealistic dependence behavior. Moreover, note that the choice of a normal distribution may appear arbitrary, but it was suggested by \cite{HT04} as a convenient choice that ensures valid inference. Indeed, under mild assumptions, even if the normal distribution is misspecified, the parameters are still guaranteed to be consistent \citep{Crowder2001}.

Although the parameters in \eqref{eq:ht} have their own individual interpretations, sometimes it can be challenging to draw conclusions on the real behavior of the conditional extremal dependence. Thus, the more preferable approach is to analyze them jointly through a functional involving them all together. To do so, we exploit the stochastic representation in \eqref{eq:w}, but we relax the normal assumption by estimating an upper conditional quantile semi-parametrically. To be more specific, we here estimate the $0.975$-quantile of $Y_{dt}$ given that $Y_{1t}$ equals its own $0.975$-quantile, by suitably combining the estimated H\&T model parameters with the empirical $0.975$-quantile of model residuals (obtained after standardizing the data by subtracting the fitted mean and dividing by the fitted standard deviation in \eqref{eq:w}). The distribution of this estimated conditional quantile is assessed using a block bootstrap procedure detailed below in Section \ref{chap:diag}.

While the parameters may be estimated separately for each channel from \eqref{eq:loglik_channel}, it is also possible to estimate them jointly using a penalized likelihood enforcing a similar degree of smoothness for estimated dependence parameters $\alpha_{db}$ across time blocks, which stabilizes their fluctuations and reduces uncertainty. The joint negative penalized log-likelihood is
$$
\tilde{\ell}_{\tilde{\tau}, \tilde{\lambda}}(\tilde{\bm{\theta}}) = \sum\limits_{d=2}^{D}\tilde{\ell}_{\tilde{\tau},d}(\tilde{\bm{\theta}}) + \tilde{\lambda} \sum\limits_{d=2}^{D}\left\{\dfrac{1}{B}\sum_{b=1}^{B}\alpha_{db}^{2} - \left(\dfrac{1}{B}\sum_{b=1}^{B}\alpha_{db}\right)^2 \right\},
$$
where $\tilde{\lambda} > 0$ is the overall roughness penalty parameter.

Alternatively, suppose a single (joint) penalization parameter imposes too severe shrinkage in the extent to which the extremal dependence varies. In this case, one may allow each channel to have its roughness penalty parameter, $\tilde{\lambda}_{d}$, performing $D-1$ model fits in parallel.

% ---- SUBSECTION ---- %
\subsection{Extremal Spectral Analysis}\label{chap:spectral}

One major novelty of our proposed Conex-Connect approach is that we apply it not only to the original time series (i.e., from a time-domain perspective) but also to the time series ``dissected'' into several frequency bands (i.e., from a frequency-domain perspective). This reveals hidden features of the extremal dependence structure between channels that only affect certain frequency bands but may not be visible by looking at the entire time series. Most of the current work on spectral analysis of electrophysiological data focus on the spectral estimation and its association with behavioral measures and brain states. None of these examine the downstream effect of extremal dependence in a reference channel on the entire network \citep{krafty11, krafty13, fiecas16, krafty17, ombao18_neuroimage, scheffler20}. Similarly, none of the work on conditional extreme-value theory (e.g., \cite{HT04}, or \cite{Keef13}) examine extremal dependence from a spectral perspective. Here, fill this gap and study the conditional extremal dependence in a brain network from a frequency-domain perspective.

\begin{figure}[!htp]
\centerline{
\includegraphics[width=.95\textwidth]{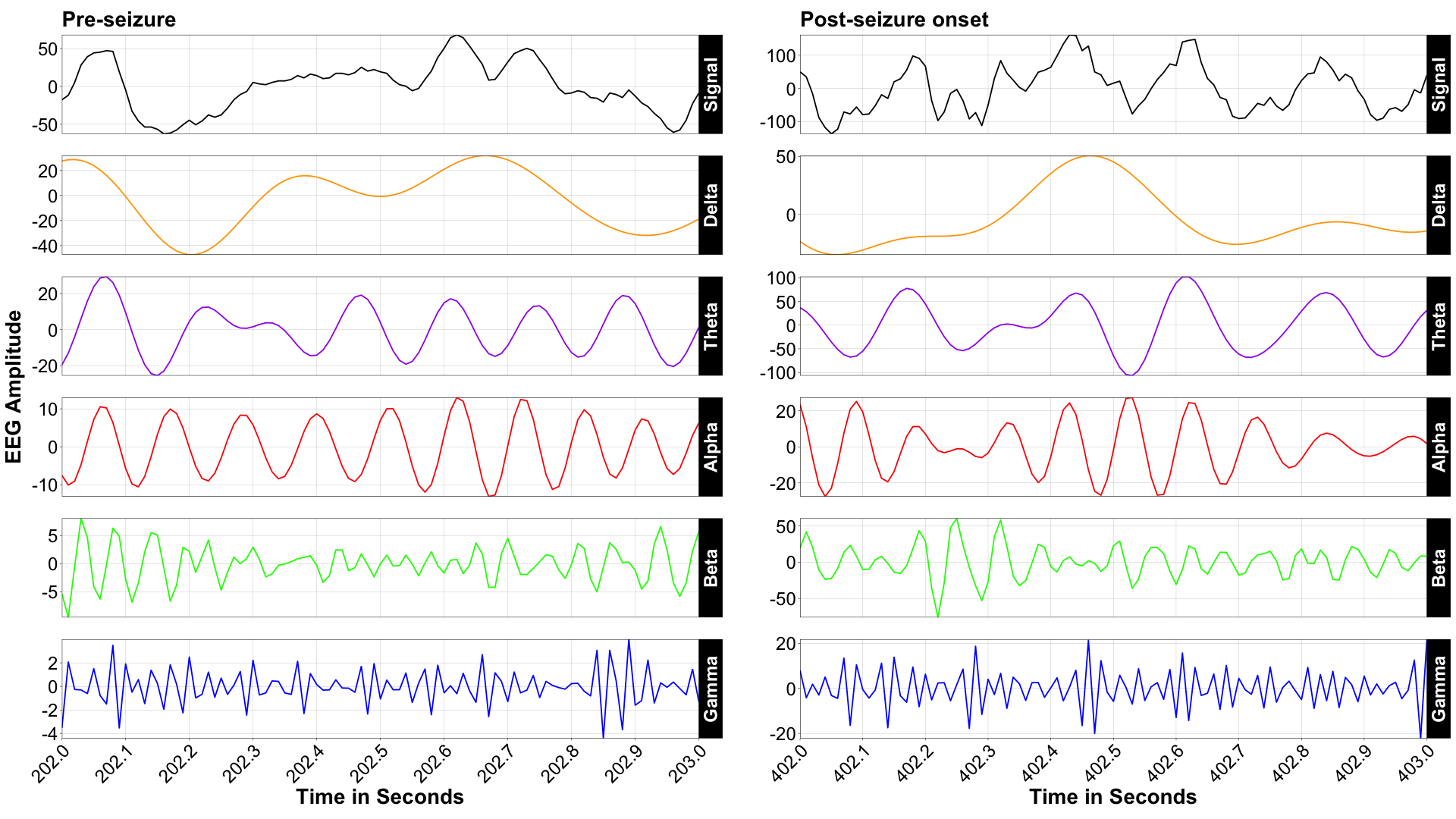}
}
\caption{Spectral decomposition of channel T3 signal during 1 second of both pre- and post-seizure onset phases from the EEG recording of a patient diagnosed with left temporal lobe epilepsy. The frequency bands are: Delta-band ($\Omega_1$: 1--4 Hz), Theta-band ($\Omega_2$:  4--8 Hz), Alpha-band ($\Omega_3$: 8--12 Hz), Beta-band ($\Omega_4$: 12--30 Hz), and 
Gamma-band ($\Omega_5$: 30--50 Hz).}\label{fig:spectral}
\end{figure}

EEGs are zero-mean signals that can be expressed as a mixture of frequencies oscillating at the following standard frequency bands: Delta-band ($\Omega_1$: 1--4 Hz), Theta-band ($\Omega_2$:  4--8 Hz), Alpha-band ($\Omega_3$: 8--12 Hz), Beta-band ($\Omega_4$: 12--30 Hz), and  Gamma-band ($\Omega_5$: 30--50 Hz); assuming a sampling rate of $100$ Hz \citep{Ombao_book, nunez07}. The spectral decomposition of channel T3 is displayed in Figure \ref{fig:spectral} for both the pre- and post-seizure onset phases. 

Following \cite{ombao2008} and \cite {gao2020}, two EEG signals, $Y_{1}(t)$ and $Y_{2}(t)$, may be decomposed into the 5 rhythms as $Y_{1}(t) \approx \sum_{k=1}^{5} Y_{1, \Omega_k}(t)$ and $Y_2(t) \approx \sum_{k=1}^{5} Y_{2, \Omega_k}(t)$, where $Y_{d, \Omega_k}(t)$ denotes the $\Omega_k$-waveform in channel $Y_{d}(t)$. In practice these rhythms are obtained by applying a linear filter such as the Butterworth filter (see \cite{cohenX14}). One measure of dependence between two channels is coherence, which is essentially the frequency band-specific squared correlation between a pair of rhythms, i.e., 
$$
\mbox{Coherence}_{\Omega_{k}}(t) = \max\limits_{\ell}{\left\lvert \mbox{cor}\left(Y_{1, \Omega_k}(t), Y_{2, \Omega_k}(t+\ell)\right)\right\rvert^2}. 
$$

Another major contribution of this paper is a novel measure of dependence based on the extremal behavior of the oscillations of brain channels. We propose a measure that examines the impact of unusually large frequency band amplitudes in the T3 channel (the reference channel projecting from the seizure foci on the temporal lobe) on the other channels. In order to study the effect of the extremal behavior in the $\Omega_k$-waveform, we study the conditional distribution of $\vert Y_{d,\Omega_i}(t) \vert$ given $\vert Y_{1,\Omega_j}(t) \vert > \tau$, $i,j=1, \ldots, 5$. The new measure produces new interesting results that give us deeper insights into the highly non-linear interactions and dependence between channels during an epileptic seizure event.

% ---- SUBSECTION ---- %
\subsection{Cross-validation, bootstrapping, uncertainty and diagnostics}\label{chap:diag}

This section gives details on both the block bootstrap and walk-forward cross-validation procedures designed to assess the uncertainty of estimated parameters and to select the optimal roughness penalty parameters respectively, while handling the auto-correlation of the time series.

\begin{figure}[!htb]
\centerline{
\includegraphics[width=.75\textwidth]{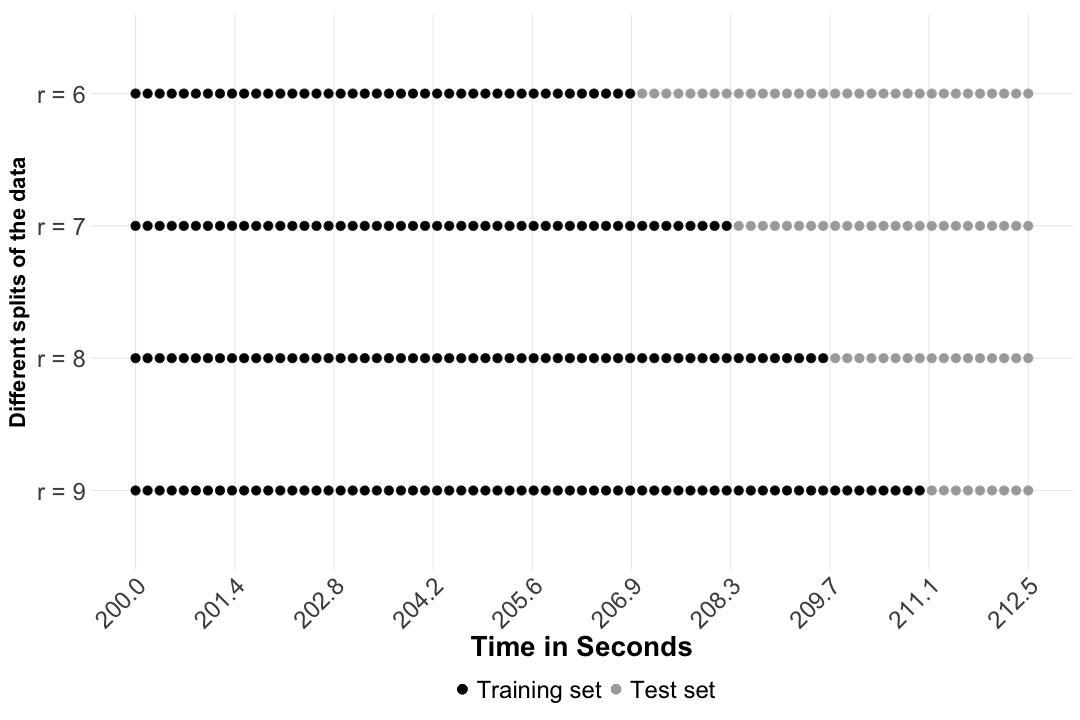}
}
\caption{Schematic for a 10-fold walk-forward cross-validation procedure for the auto-correlated data within time block 1 of the pre-seizure phase of the EEG application. The blue dots represent the training sets. The red dots are the test sets.}
\label{fig:cv_scheme}
\end{figure}

In both the marginal and the H\&T models, we need to select the optimal value of the roughness penalty parameter, $\lambda >0$, in an objective manner. In both cases, for each channel and within each time block, we rely on a $10$-fold walk-forward cross-validation procedure \citep{hyndman} since the data are auto-correlated across time. For this procedure, we divide each time block $b$ into $k = 10$ adjacent non-overlapping folds. Then, we consider the first $r$ folds as the training set and the last $k-r$ folds as the test set. For a grid of $\lambda$ values and for each split configuration, we then fit the model to the training set. We use the parameter estimates to obtain the value of the log-likelihood function on the test set, and we finally add up these log-likelihood values of each split configuration to get an overall score. The value of $\lambda$ that maximizes the overall score is taken as the optimal value for the penalization parameter. Figure \ref{fig:cv_scheme} shows a schematic illustration of this cross-validation procedure.

To assess the uncertainty of parameter estimates, we employ a block bootstrap procedure which preserves local temporal dependence in the data. As described in \cite{lahiri}, we resample entire blocks of consecutive observations (rather than single observations) keeping the correspondence between the conditioning variable $Y_1$ and the associated variables $Y_2, \ldots, Y_D$, and then refit the marginal and H\&T dependence models. From the bootstrap samples, we compute confidence intervals for the parameters that evolve with time and histograms for those that are constant over time. Notice that the bootstrap blocks differ from the $B$ time blocks used in the PPC model formulation. 

Specifically, we split each of these $B$ time blocks into $C$ non-overlapping (sub-)blocks, of size $s$, which are resampled with replacement to generate $M$ bootstrap samples. The block size $s$ is determined with the aid of the sample autocorrelation function to make sure it is large enough to keep the overall temporal dependence structure.

Simultaneously, for each bootstrap sample, the non-exceedance probability for both mar\-ginal and H\&T models is also randomly sampled from a uniform distribution given a range of reasonable values to account for thresholding uncertainty: $\tau_d \in \mathcal{T}_d \subset (0, 1)$, $d\in\{1, \ldots, D\}$, and $\tilde{\tau} \in \tilde{\mathcal{T}} = \subset (0, 1)$. In this way, one may produce diagnostic plots, such as parameter stability plots, to select the best threshold for each channel. Also, from the bootstrap samples, quantile-quantile (Q-Q) plots and residual plots can be drawn to evaluate the goodness-of-fit.

% ---- ------- ---- % ------- % ---- ------- ---- %
% ---- SECTION ---- % SECTION % ---- SECTION ---- %
% ---- ------- ---- % ------- % ---- ------- ---- %
\section{Extremal Spectral Analysis of Seizure EEG Data}\label{chap:application}

% ---- SUBSECTION ---- %
\subsection{Data and Statistical Analysis}

We examine the changes in connectivity (or dependence between channels) using the proposed Conex-Connect method. In this paper, the scalp EEG recording is from a female patient diagnosed with left temporal lobe epilepsy. The EEG data is recorded from a $10$--$20$ system, and $D=19$ channels are available for analysis. Figure \ref{fig:eeg_scalp} shows the electrode placement and EEG traces for some channels. In this specific case, the physician knew in advance that the seizure onset would be on the left temporal lobe region, which would be captured by the recording at the T3 channel ($d = 1$). Thus, T3 is set to be the reference channel (i.e., the conditioning variable in the H\&T model). The remaining 18 channels are treated as the associated variables ($d = 2, \ldots, 19$). The record has a duration of $500$ seconds collected at the sampling rate of $100$ Hz, thus there are $50,000$ time points per channel.

The first stage of the Conex-Connect method is to model the marginal extremes for each channel $d$ independently. Each channel $d$ is split into pre- and post-seizure onset phases at time $t = 350$ seconds and thus $T = 15,000$ points for each of the post-seizure onset and pre-seizure phase. Each phase is segmented into $B = 12$ time blocks of $1,250$ data points each. With this setup, the GP distribution is fitted jointly for all time blocks, specifying block-specific scale parameters and a constant shape parameter, as indicated in Section \ref{chap:marginals}, and the fitted model is then used to transform the data to the standard Laplace scale. Moreover, to assess estimation uncertainty, $M=500$ block bootstrap samples are generated within each time block $b=1,\ldots,B=12$, with bootstrap block size $s = 25$ data points. For all channels $d = 1,\ldots, D$, a common non-exceedance probability interval, $\mathcal{T}_{d} = (0.90, 0.95)$, is used. The results and diagnostics of this marginal estimation stage are provided in the Supplementary Material.

The second stage of the Conex-Connect method is to model the conditional extremal dependence of brain connectivity. The strength of the relationship between the associated variables and large values of the conditioning variable, T3, is estimated. After transforming the data from all channels into the standard Laplace scale, we use the H\&T model where the non-exceedance probability interval for the reference channel is chosen to be $\tilde{\mathcal{T}} = (0.88, 0.92)$, both for the pre- and post-seizure onset phases. Estimation, uncertainty assessment, and goodness-of-fit details are thoroughly reported in the Supplementary Material. For the sake of brevity, we here only present and discuss the results from the associated frontal channels, namely the left frontal F7 and right frontal F8 (see Section \ref{chap:results}), while dashboards for the other channels can be found in the Supplementary Material for completeness.

% ---- SUBSECTION ---- % 
\subsection{Results and Discussion}\label{chap:results}

\begin{figure}[!htp]
\centerline{
\includegraphics[width = .665\textwidth]{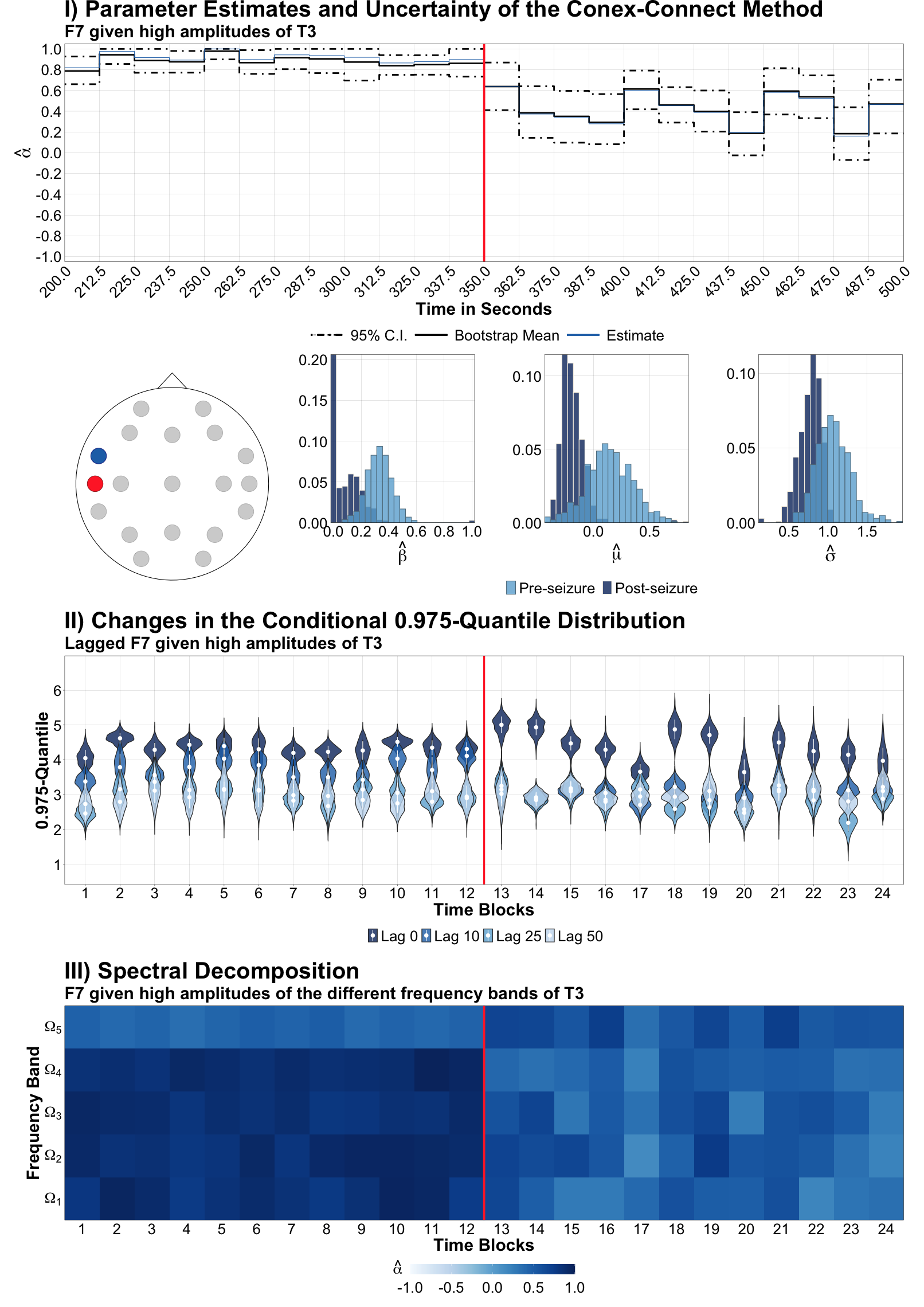}
}
\vspace*{-.5cm}
\caption{A dashboard with results of the Conex-Connect method for pre- and post-seizure onset phases. Channel F7 (blue) given high values of T3 (red) are highlighted in the EEG scalp cartoon. \textbf{Panel I)} In the first line, the evolution of the estimated first-order dependence parameter $\alpha_{db}$ (solid blue line) through time with its bootstrap mean (solid black line) and 95\% confidence bands (dashed lines). In the second line, the histograms of the bootstrap estimates for scale exponent parameter $\beta_d$, and for residual mean $\mu_d$ and scale $\sigma_d$. \textbf{Panel II)} Bootstrap violin plots for the $0.975$-quantile of the conditional distribution of F7 given that T3 reaches its own $0.975$-quantile. Different colors represent estimates of the conditional quantile for different lag values of the associate channel F7. \textbf{Panel III)} Effect of the different frequency bands ($\Omega_1$: 1--4, $\Omega_2$: 4--8, $\Omega_3$: 8--12, $\Omega_4$: 12--30, and $\Omega_5$: 30--50 Hz) on the first-order dependence parameter. Darker blue pixels indicate higher dependence.}
\label{fig:ht_f7}
\end{figure}

Figure \ref{fig:ht_f7} displays a dashboard with the results of the Conex-Connect method for channel F7 given high values of the reference channel T3, both for pre- and post-seizure onset phases. Panel I) provides the individual behavior of the H\&T parameter estimates: $\{\hat{\alpha}_{2,b}\}_{b=1}^{B}$, $\hat{\beta}_2$, $\hat{\mu}_2$, and $\hat{\sigma}_2$. Here, the subscript $d = 2$ refers to channel F7 and $b \in \{1, \ldots, 24\}$ denotes the time block index. The 95\% confidence bands for $\hat{\alpha}_{2,b}$ and the histograms for $\hat{\beta}_2$, $\hat{\mu}_2$, and $\hat{\sigma}_2$ are drawn out from the block bootstrap procedure; see Section \ref{chap:diag}. In the pre-seizure phase, the estimates of the first-order dependence parameter, $\hat{\alpha}_{2,b}$, $b=1, \ldots, 12$, are very high and stable, close to $1$, indicating a strong positive linear relationship between F7 and large values of T3. This makes sense as both channels F7 and T3 are located on the left side of the brain. Note that the confidence bands are narrow, suggesting a high level of certainty concerning this dependence structure. The histogram of $\hat{\beta}_2$ for the pre-seizure phase is centered around $0.4$ with practically no values close to $0$, indicating asymptotic independence, yet with strong sub-asymptotic dependence (i.e., at finite levels). In the post-seizure phase, we can see a sudden change immediately after the seizure, with $\hat{\alpha}_{2,13}$ being much smaller than $\hat{\alpha}_{2,12}$, indicating weaker extremal dependence. Also, the behavior of $\hat{\alpha}_{2,b}$, $b = 13, \ldots, 24$, are much more erratic with wider confidence bands, suggesting a higher level of uncertainty after the seizure onset. This is interesting because it reveals the more chaotic nature of the seizure process. For some blocks, the estimates are close to $0.2$. Thus, there is a weakening in the linear relationship between F7 and large values of T3, though it still remains positive. In terms of $\beta_2$, there is a shift to the left in the histogram with a high concentration around $0$, pointing to a possible change towards independence.

\begin{figure}[!htp]
\centerline{
\includegraphics[width = .665\textwidth]{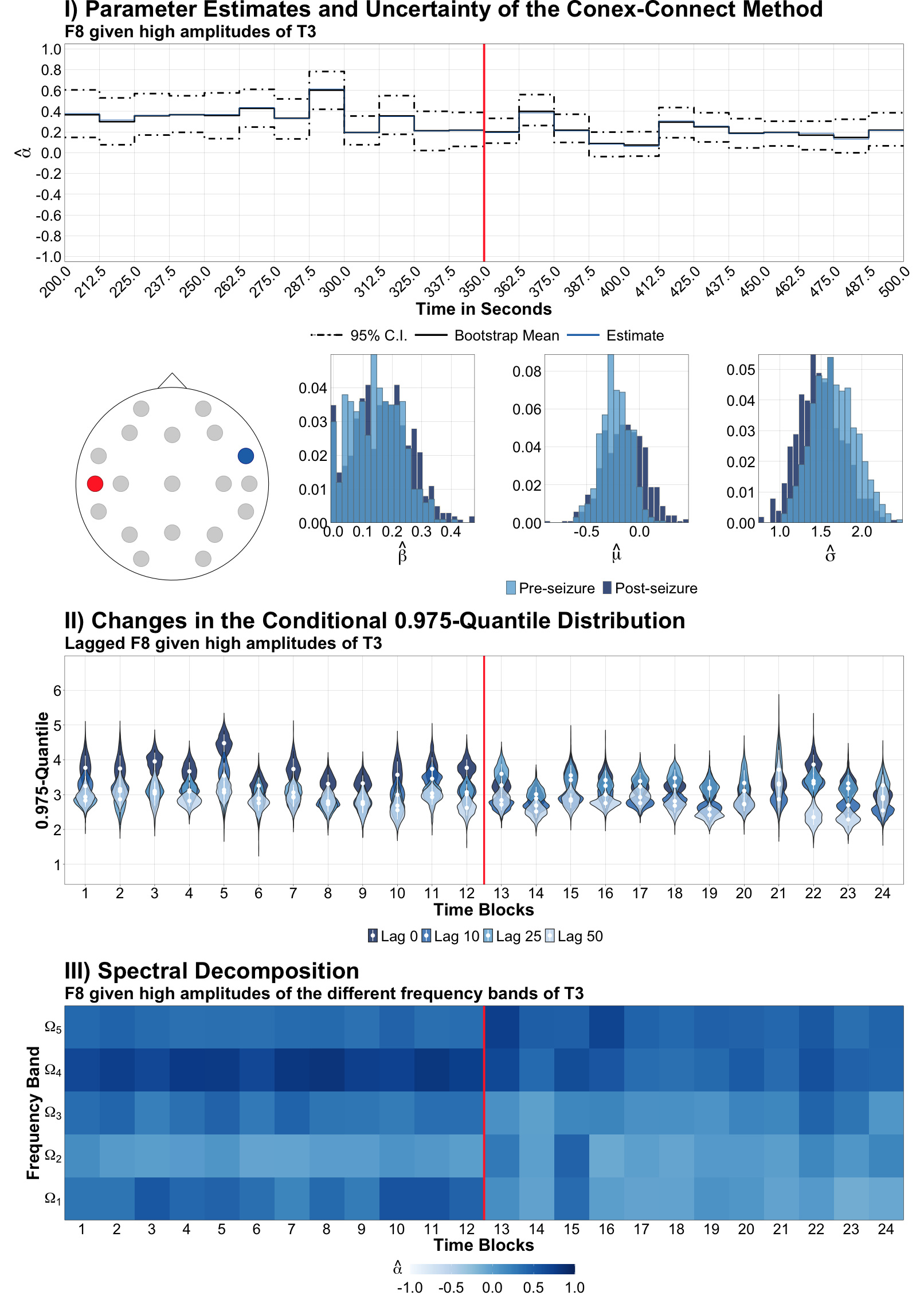}
}
\vspace*{-.5cm}
\caption{A dashboard with results of the Conex-Connect method for pre- and post-seizure onset phases. Channel F8 (blue) given high values of T3 (red) are highlighted in the EEG scalp cartoon. \textbf{Panel I)} In the first line, the evolution of the estimated first-order dependence parameter $\alpha_{db}$ (solid blue line) through time with its bootstrap mean (solid black line) and 95\% confidence bands (dashed lines). In the second line, the histograms of the bootstrap estimates for scale exponent parameter $\beta_d$, and for residual mean $\mu_d$ and scale $\sigma_d$. \textbf{Panel II)} Bootstrap violin plots for the $0.975$-quantile of the conditional distribution of F8 given that T3 reaches its own $0.975$-quantile. Different colors represent estimates of the conditional quantile for different lag values of the associate channel F8. \textbf{Panel III)} Effect of the different frequency bands ($\Omega_1$: 1--4, $\Omega_2$: 4--8, $\Omega_3$: 8--12, $\Omega_4$: 12--30, and $\Omega_5$: 30--50 Hz) on the first-order dependence parameter. Darker blue pixels indicate higher dependence.}
\label{fig:ht_f8}
\end{figure}

By contrasting the findings above to panel I) of Figure \ref{fig:ht_f8}, we can see that a different story prevails on the right side of the brain. For the right frontal channel F8 ($d=3$), given high values of the reference channel T3, it appears that seizure does not affect the extremal dependence structure as much, since there is practically no discernible change in the estimates $\hat{\alpha}_{3, b}$ before and after the seizure. Also, in the pre-seizure phase, the linear relationship between F8 and large values of T3 is weaker when compared to F7. This is partly because F7 is closer to the seizure focus location, which is around the reference channel T3. In addition, the histograms for $\hat{\beta}_3$, $\hat{\mu}_3$, and $\hat{\sigma}_3$, both for pre- and post-seizure onset phases, are relatively stable, indicating the absence of changes in the second-order extremal dependence characteristics.

Back to Figure \ref{fig:ht_f7}, panel II) displays the time evolution of the conditional distribution of the conditional 0.975-quantile in \eqref{eq:w} given that T3 reaches its own 0.975-quantile, estimated semi-parametrically as explained in Section \ref{chap:dependence}. In this plot, we can analyze the parameters of the H\&T model jointly. Also, with the different violin plots, we investigate how the effect of large values of T3 at time $t$ impacts channel F7 at time $t+\ell$, for time lags $\ell = 0$, $0.10$, $0.25$, and $0.50$ seconds. The Conex-Connect method shows that the extremal dependence is stronger at lag $\ell=0$ and weakens for other lags, indicating a stronger contemporaneous extremal dependence than lagged extremal dependence. Moreover, when contrasting lag $\ell=0$ to the other lags, the discrepancy between the violins, both in terms of medians and shapes, is more pronounced after the seizure onset. This suggests that seizure has an impact on the conditional extremal dependence of the brain network. The same panel of Figure \ref{fig:ht_f8}, in the case of channel F8, the opposite side from the seizure focus, shows that the extremal dependence structure at lag $\ell=0$ is indistinguishable from those of higher lags, and at a lower level overall than for channel F7. This denotes less synchrony (in the extremal dependence) between the channel corresponding to the seizure foci (reference channel) and the channel on the contra-lateral side of the foci. 

The Conex-Connect method produces interesting results regarding how the extreme values of the different oscillations of the reference channel T3 impact brain connectivity. Figure \ref{fig:ht_f7}, panel III), shows the estimated first-order dependence coefficient $\hat{\alpha}_{2,b}$ ($b=1,\ldots,B$) between the (absolute) $\Omega_k$-waveform in channel F7 given high (absolute) amplitudes of the same waveform in channel T3. Values of $\hat{\alpha}_{2,b}$ closer to 1 are darker. In the pre-seizure phase, the extreme values in the high-frequency Gamma-band exhibit the lowest level of extremal dependence. This seems to be consistent across the entire pre-seizure phase. However, the dependence pattern changes in the post-seizure phase. First, the extremes of Gamma-band from T3 shows the highest level of extremal dependence with the Gamma-band from F7. This is quite interesting because sudden outbursts of high-frequency oscillations typically characterize seizure onset as shown by \cite{medvedev11}. Moreover, since the post-seizure onset is typically non-stationary, we see that this dependence structure also evolves over time blocks. In the right side of the brain, Figure \ref{fig:ht_f8}, panel III), shows that before the seizure occurs, the higher values of the mid-frequency Beta-band lead the changes in the extremal dependence structure. In the post-seizure phase, we notice that, immediately after the seizure onset, the high-frequency Gamma-band becomes more prominent, similar to the left side of the brain (channel F7). 
 
\begin{figure}[!thp]
\centerline{
\includegraphics[width=.90\textwidth]{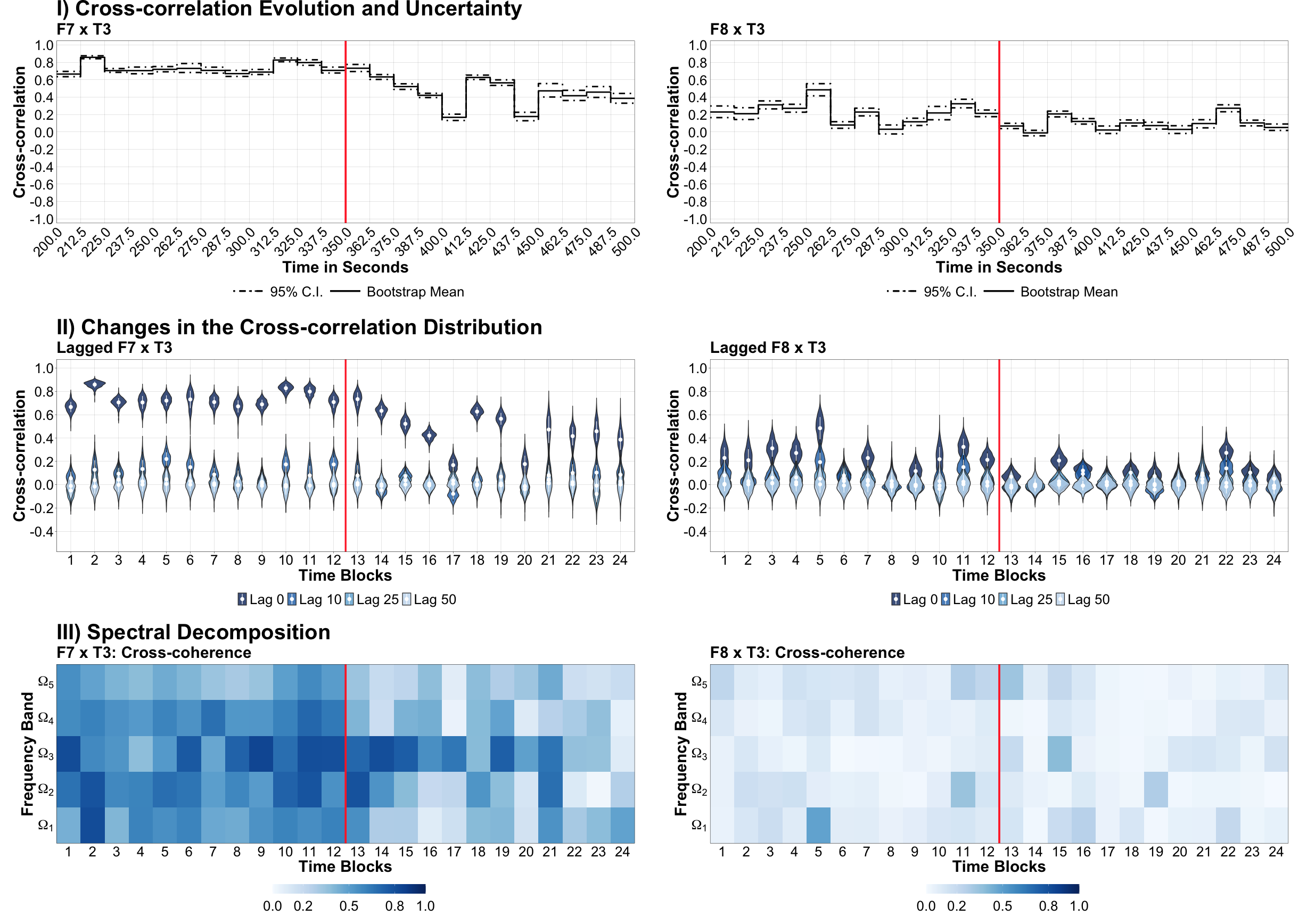}
}
\vspace*{-.5cm}
\caption{A dashboard with results of the classical analysis for pre- and post-seizure onset phases. Left column displays results for channel F7 (brain left size) while the right column show the results for F8 (brain right side). \textbf{Panel I)} Evolution of the cross-correlation estimates (black solid line), along with its confidence bands (dashed lines), between channels F7 and T3 (reference channel) and between channels F8 and T3. \textbf{Panel II)} Violin plots to access the changes in the distribution of the cross-correlation both over time and for different lag values of the associated channels. \textbf{Panel III)} Effect of the different frequency bands on the cross-coherence. Darker blue pixels indicates higher values of cross-coherence. For all panels, uncertainty is obtained trough 500 bootstrap samples.}
\label{fig:classical_f7xf8}
\end{figure}

Figure \ref{fig:classical_f7xf8}, presents a dashboard with the results of classical methods based on cross-correlation and cross-coherence for comparison. The left column shows the results for F7, while the right column shows the results for F8. In all panels, we display both pre- and post-seizure onset phases. Panel I) displays the evolution of the cross-correlation over time blocks. Panel II) shows how the cross-correlation between T3 at time $t$ evolves when computed for future values of channel F7 and F8 at time $t+\ell$, for time lags $\ell = 0$, $0.10$, $0.25$, and $0.50$ seconds. Finally, panel III) exhibits the impact of high values of the different frequency bands of T3 in its cross-coherence with F7 and F8.

\begin{figure}[!htp]
\begin{subfigure}[c]{\textwidth}
  \centering
  % include first image
  \includegraphics[width=.85\linewidth]{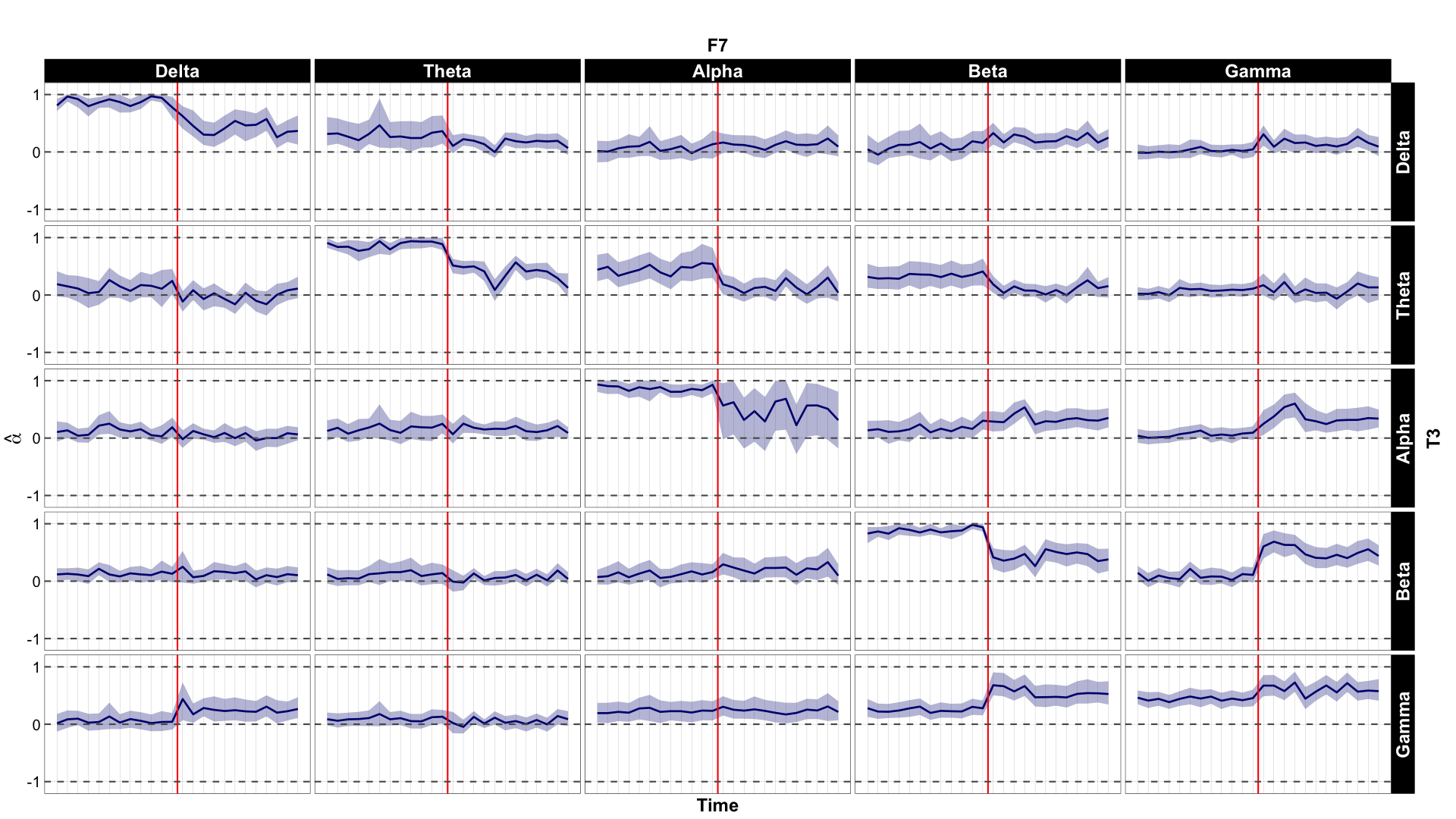}  
\end{subfigure}
\vskip\baselineskip\vspace*{-.75cm}
\begin{subfigure}[c]{\textwidth}
  \centering
  % include second image
  \includegraphics[width=.85\linewidth]{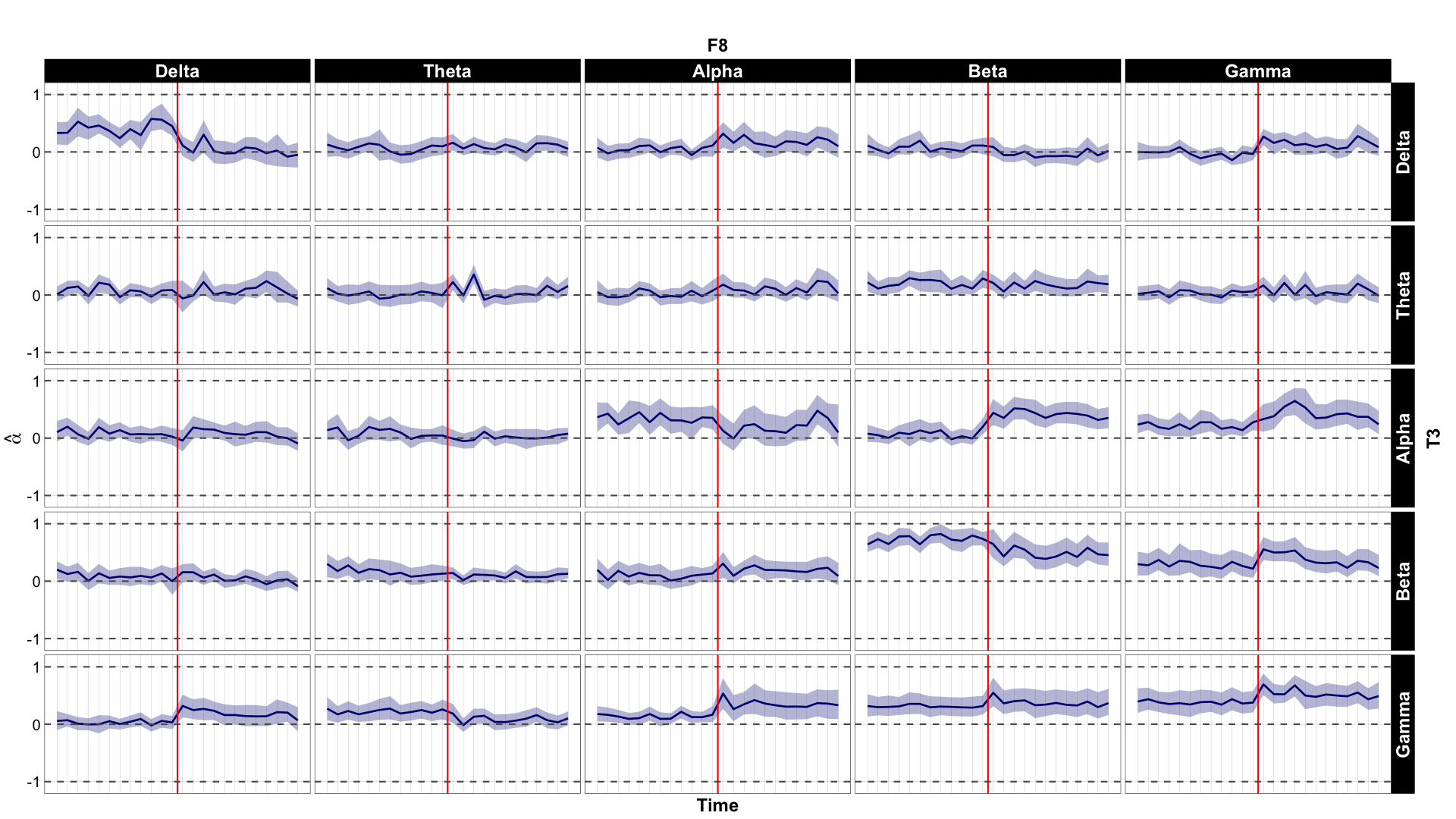}  
\end{subfigure}
\caption{Effect of the different frequency bands of the reference channel T3 on the different frequency bands of the associated channels, F7 and F8, pre- and post-seizure onset phases. The solid blue line represents the evolution of the estimated first-order dependence parameter $\alpha$ through time with its bootstrap 95\% confidence bands (shaded area). The vertical red line is the seizure onset. The frequency bands are: Delta-band ($\Omega_1$: 1--4 Hz), Theta-band ($\Omega_2$:  4--8 Hz), Alpha-band ($\Omega_3$: 8--12 Hz), Beta-band ($\Omega_4$: 12--30 Hz), and Gamma-band ($\Omega_5$: 30--50 Hz).}
\label{fig:5x5_plots}
\end{figure}

When we contrast the results of our method Conex-Connect, based on extreme-value theory with classical results, we notice that the first-order dependence parameters of our method have a similar overall temporal pattern as that of the classical cross-correlation, both for F7 and F8, pre- and post-seizure, as can be seen when comparing Figures \ref{fig:ht_f7}, \ref{fig:ht_f8}, and \ref{fig:classical_f7xf8}. In addition, comparing the changes in the conditional 0.975-quantile with the changes in the cross-correlation (panel II), Figure \ref{fig:classical_f7xf8}), we observe a similar pattern across time blocks both for pre- and post-seizure onset phases. However, regarding correlation, we notice a more striking discrepancy between contemporaneous dependence (lag $\ell=0$) against other higher lags. These similarities suggest either (a.) that the conditional extremal dependence dominates the global dependence (as measured by correlation) or (b.) that the phenomenon that we see in the joint tail is similar to other less extreme quantiles of the distribution. Regarding the spectral decomposition, our method shows that for both sides of the brain, immediately after the seizure, the high-frequency Gamma-band becomes the most relevant frequency in explaining the conditional extremal dependence. This finding does not agree with the results in terms of the classical coherence; see Figure \ref{fig:classical_f7xf8}, panel III). We further investigate the conditional extremal dependence in terms of frequency oscillations by decomposing the associated channels in their canonical frequency bands. We refit the model for all pairs of $\Omega_k$-waveforms. Figure \ref{fig:5x5_plots} displays the impact of extreme values of the different frequency bands of the reference channel T3 on the different frequency bands of the associated channels F7 and F8. Note that the heatmap in panel III) of Figures \ref{fig:ht_f7} and \ref{fig:ht_f8} corresponds to the diagonals of Figure \ref{fig:5x5_plots}. Here, beyond the previous findings in terms of seizure phases and sides of the brain, we notice that the medium-frequency bands, Beta and Alpha, of the reference channel T3, impact the first-order dependence parameter estimates, mainly on the Gamma-band of the associated channels. This may indicate that the Gamma-band can be used for feature engineering to improve the performance of machine learning algorithms for epilepsy detection.

% ---- ------- ---- % ------- % ---- ------- ---- %
% ---- SECTION ---- % SECTION % ---- SECTION ---- %
% ---- ------- ---- % ------- % ---- ------- ---- %
\section{Conclusion}\label{chap:conclusion}

In this paper, we present the novel Conex-Connect method, the first extreme-value model-based approach to learn patterns in the extremal dependence during periods of high volatility in brain signals. The method extends the \cite{HT04} model to capture time-varying extremal dependence features, both from time-domain and frequency-domain perspectives, while adequately assessing estimation uncertainty using a block bootstrap procedure. We here give a full characterization of the conditional extremal dependence of brain connectivity, shedding light on how the brain network responds to an epileptic seizure event. We also study the extremal brain connectivity based on the spectral decomposition of the different channels. To the best of our knowledge, our method is a pioneer in linking the association between extreme values of frequency oscillations in a reference channel with oscillations in other channels of the brain network. 

Essentially, our methodology relies on ``dissecting'' the original brain signal into various components at different frequency bands, and to study the extremal dependence structure of each frequency component separately based on distinct conditional extreme-value models. While our statistical analysis is motivated by a very concrete applied problem and our empirical results reveal salient features of brain connectivity during an epileptic seizure, it would also be interesting in future research to further investigate the theoretical link between the (conditional) extremal dependence structure of each frequency component to that of the overall signal. Specifically, we expect that the frequency component with the strongest tail dependence would dominate the overall tail dependence structure and dictate the occurrence of joint extremes in the original time series. However, this conjecture still needs to be rigorously validated and mathematically formalized. Conversely, the overall extremal dependence structure imposes certain constraints on the joint behavior of each frequency component, and it would be interesting to study this in more detail.

Our proposed approach demonstrates changes in the conditional extremal dependence of brain connectivity between pre- and post-seizure onset phases. In general, before the seizure, the dependence is notably stable for all channels (conditioning on extreme values of the T3). On the other hand, during the post-seizure phase (a period of very high volatility at T3), the dependence between channels is weaker. In general, the strength of dependence decreases at large lagged values of the associated channels when T3 is kept fixed. Also, after the seizure, the high values of the high-frequency Gamma-band are the most relevant features to explain the conditional extremal dependence of brain connectivity.

Currently, our method does not capture the underlying spatial structure of the brain. This is a complex topic because the usual Euclidean distance is not appropriate. Hence, as a future research goal, we aim to develop extreme-value-based models to deal not only with the time-varying features of brain connectivity but also with its spatio-temporal characteristics. A possibility would be to adapt the multivariate version of the H\&T model or the conditional spatial extremes model of \cite{2019arXiv191206560W} to the time-varying framework of brain signals, combined with an appropriate distance metric.

% ---- ------- ---- % ------- % ---- ------- ---- %
% ---- SECTION ---- % SECTION % ---- SECTION ---- %
% ---- ------- ---- % ------- % ---- ------- ---- %
\begingroup
\fontsize{12pt}{11pt}\selectfont
\bibliographystyle{chicago}
\bibliography{Bibliography}
\endgroup

\end{document}